\documentclass[12pt]{article}

\usepackage{amsmath, amssymb, amsfonts, amsthm} \usepackage{graphicx}
\usepackage{natbib} \usepackage{indentfirst} \usepackage{times}
\usepackage{bbm} \usepackage{lineno} \usepackage{authblk}
\usepackage{afterpage}
\usepackage{rotating}

\usepackage{lineno}
\allowdisplaybreaks

\usepackage[colorlinks, linkcolor=black,
citecolor=black,filecolor=black, urlcolor=black]{hyperref}

\newcommand{\transpose}[1]{\ensuremath{{#1}^{\!\mathrm{T}}}}

\makeatletter

\newcounter{id} \renewcommand{\theid}{I\arabic{id}}
\renewcommand{\theHid}{I\arabic{id}}

\newcommand{\c@org@eq}{} \let\c@org@eq\c@equation
\newcommand{\org@theeq}{} \let\org@theeq\theequation

\newcommand{\setid}{ \let\c@equation\c@id \let\theHequation\theHid
\let\theequation\theid}

\newcommand{\setreg}{ \let\c@equation\c@org@eq
\let\theHequation\org@theHeq \let\theequation\org@theeq}

\makeatother

\begin{document}

\title{The diversity of evolutionary dynamics on epistatic versus non-epistatic fitness landscapes
}

\author[1]{David M.~McCandlish\thanks{To whom correspondence should be
addressed. E-mail: \texttt{davidmc@sas.upenn.edu}}} \author[1]{Jakub Otwinowski} \author[1]{Joshua B.~Plotkin} \affil[1]{Department of
Biology, University of Pennsylvania, Philadelphia, PA}

\date{}

{\let\newpage\relax\maketitle}

\begin{abstract}  \normalsize

 \noindent The class of epistatic fitness landscapes is much more diverse than the class of non-epistatic landscapes, and so it stands to reason that there exist dynamical phenomena that can only be realized in the presence of epistasis. Here, we compare
evolutionary dynamics on all finite epistatic landscapes versus all finite
non-epistatic landscapes, under weak mutation.  We first analyze the mean fitness
trajectory ---that is, the time course of the expected fitness of a population.
We show that for any epistatic fitness landscape and starting genotype,
there always exists a non-epistatic fitness landscape and starting genotype that
produces the exact same mean fitness trajectory.  Thus, surprisingly, the space of mean fitness trajectories that can be realized by epistatic landscapes is no more diverse than the space of mean fitness trajectories that can be realized by non-epistatic landscapes. On
the other hand, we show that epistatic fitness landscapes can produce dynamics in the time-evolution of the variance in fitness across replicate populations and in the time-evolution of the expected number of substitutions that cannot be produced by any non-epistatic landscape.  These results on identifiability 
have implications for efforts	 to infer epistasis from the types of data often measured in experimental
populations.
\end{abstract}

\vfill

%\newpage \linenumbers
%
%
\setlength{\parindent}{.62cm} \setlength{\parskip}{2ex plus0.5ex
minus0.2ex}
\setid \setreg

\section{Introduction}

A basic problem in evolutionary biology is to understand how the structure of a
fitness landscape affects the dynamics of
adaptation~\citep{Wright32,MaynardSmith70,Kauffman93,Whitlock95,Fontana98b,Weinreich05c,Jain07,Kryazhimskiy09,deVisser14,Hartl14}.
One simple classification distinguishes non-epistatic
fitness landscapes, where the fitness effects of mutations are independent of
genetic background, from epistatic landscapes, where some mutations have
background-dependent
effects~\citep{Fisher18,Weinberger90,Phillips08,deVisser11,Szendro13,Weinreich13,deVisser14}.
Because the class of epistatic fitness landscapes is far more diverse than the
class of non-epistatic fitness landscapes, it seems intuitive that
the dynamics of adaptation possible on epistatic fitness landscapes
should be more diverse than the dynamics possible on
non-epistatic landscapes. 

Here, we rigorously test this intuition for several aspects of the dynamics of
adaptation, such as the time-evolution of the expected degree of adaptation, the
expected number of substitutions, and the variance in fitness across replicate
populations. Assuming that mutation is weak, i.e.~that each new mutation is lost
or goes to fixation before the next new mutation enters the
population~\citep[see][for a review]{McCandlish13f}, we characterize the possible
dynamics of these descriptors of adaptation, across the entire class of finite
non-epistatic fitness landscapes.  We then ask whether there exist epistatic
landscapes for which these descriptors behave in ways that are impossible in the
absence of epistasis. Such an analysis provides insight into the role of epistasis
in adaptation by identifying dynamical phenomena that are possible only when
epistasis is present.

Our most important result concerns how the expected fitness of a population
changes over time. In particular, we consider an ensemble of replicate populations
that begin fixed for some genotype on a fitness landscape. We then ask, for any
given time in the future, what the mean fitness across the ensemble of populations
will be. The time evolution of this expected fitness is known as the ``mean
fitness trajectory''~\citep{Kryazhimskiy09}. Surprisingly, we prove that, for any
finite epistatic fitness landscape and choice of starting genotype, one can always
construct a non-epistatic fitness landscape that produces the exact same mean
fitness trajectory as the epistatic landscape produces. Thus, at least for the
expected degree of adaptation, the dynamics possible on epistatic landscapes are
no more diverse than those possible on non-epistatic landscapes.

\begin{figure} \center \includegraphics[width=10cm]{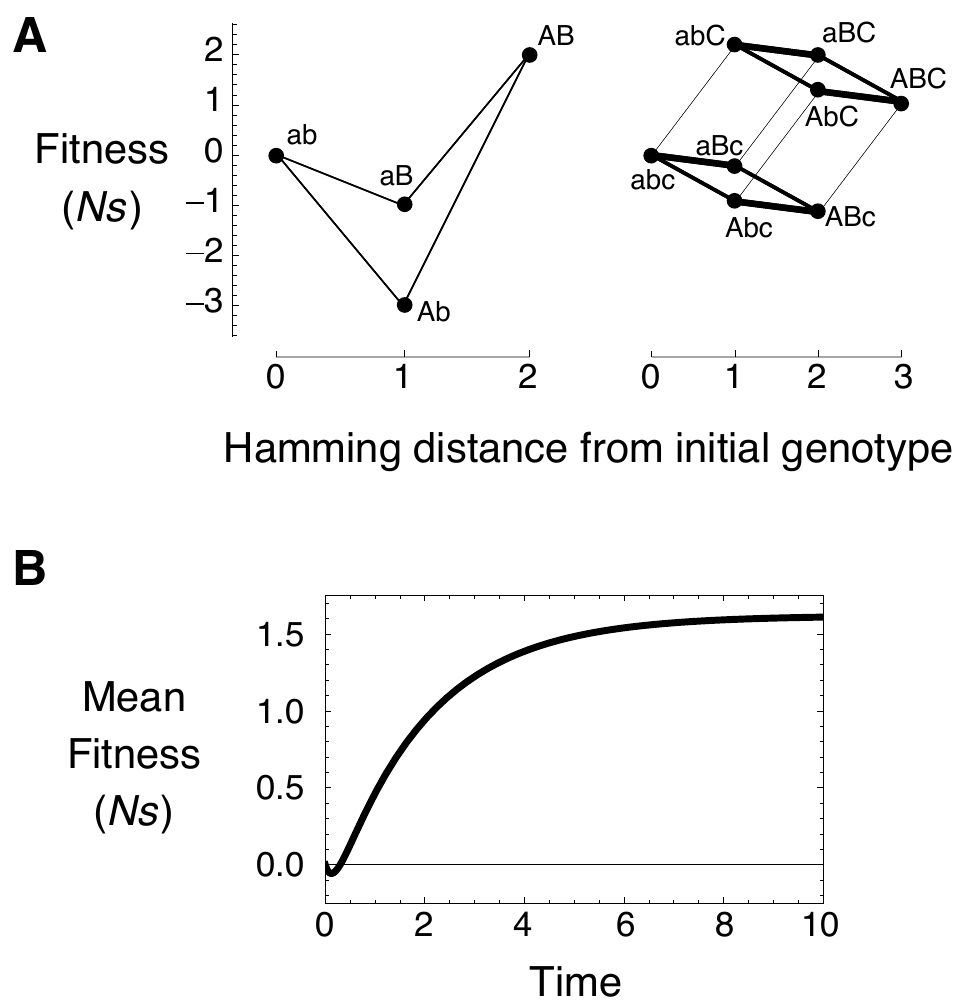}
\caption{
(A) A fitness landscape with reciprocal sign epistasis and a
non-epistatic fitness landscape that both produce the exact same mean
fitness trajectory (B) when the population begins at the left-most genotype (ab or abc). Edges are
mutations and points are genotypes. The height of a genotype indicates
its scaled selection coefficient ($Ns$) relative to the initial (left-most) genotype. Edge thickness is proportional
to mutation rate. The bottom panel shows expected fitness as a function
of time (the mean fitness trajectory), where time is measured in terms of
expected substitutions at a neutral locus with a unit mutation rate.
Notice that the trajectory is quite complex; in particular, it
decreases slightly at short times before increasing to its asymptotic
value at long times.
\vspace{1.5cm}
}
\label{fig:illustration} \end{figure}

To illustrate this general result in a specific case,
Figure~\ref{fig:illustration}A shows an epistatic and a non-epistatic fitness
landscape that produce the same mean fitness trajectory
(Figure~\ref{fig:illustration}B) for a population that starts at the left-most
genotype (ab or abc).  Note that the first fitness landscape in
Figure~\ref{fig:illustration}A is not merely epistatic, but in fact exhibits
reciprocal sign epistasis~\citep{Weinreich05c}, and the resulting mean fitness
trajectory decreases initially (due to deleterious fixations into the
fitness valley) before increasing towards its asymptotic value, as populations
become more likely to have crossed the fitness valley. It may seem surprising that
one can construct a non-epistatic landscape that produces the same, complex
dynamics. And, yet, our main result says that constructing such a landscape is
always possible.

Our main result shows that the set of realizable mean fitness trajectories for
epistatic landscapes is no more diverse than the set of mean fitness trajectories for
non-epistatic landscapes, despite the fact that, e.g.~epistatic fitness landscapes
can have multiple fitness peaks while non-epistatic landscapes are always
single-peaked. However, the presence of epistasis does increase the diversity of
possible dynamics for several more subtle descriptors of the adaptive process. For
instance, if we again consider an ensemble of replicate populations evolving from
the same starting genotype, we can study the variance in fitness across these
populations as a function of time. It turns out that there exist epistatic fitness
landscapes whose ``variance trajectories'' have features that cannot be achieved on
any non-epistatic landscape.  In particular, a variance trajectory that is
accelerating at short times can occur only on an epistatic fitness landscape.  A
similar result holds for the expected number of substitutions that accrue over
time, i.e.~the mean substitution trajectory~\citep{Kryazhimskiy09}: for any
non-epistatic fitness landscape the expected substitution rate can never be less
than half the equilibrium substitution rate, whereas some epistatic fitness
lanscapes violate this condition and are expected to produce a broader range of
substitution rates.

Our emphasis here is on the theoretical question of the diversity of evolutionary dynamics with or
without epistasis, rather than on developing practical tests for epistasis.
Nevertheless, our results show that epistasis is in principle identifiable from
some statistics, but not from others. Our results therefore have implications for
attempts to infer the presence and form of epistasis from empirical observations
of evolution in replicate laboratory
populations~\citep[e.g.,][]{Lenski91,Lenski94,Kryazhimskiy09,Wiser13}, an approach
that has recently been the subject of some controversy~\citep{Frank14,Good14}.  We
return to this problem of inferring epistasis from observed evolutionary dynamics
in the Discussion.

\section{Results}

\subsection{Population-genetic model}

We consider the space of all possible fitness landscapes with a finite number of
bi-allelic sites. A genotype is defined by the state at all of its sites, and the
fitness of the $i$-th genotype is denoted $\mathbf{F}(i)$.  We will work in scaled
Malthusian fitness (i.e.~the fitness of a genotype is equal to the logarithm of
the standard, Wrightean fitness times the population size) so that the scaled
selection coefficient of genotype $j$ relative to genotype $i$ is given by
$\mathbf{F}(j)-\mathbf{F}(i)$.

Our main population-genetic assumption is that mutation is weak, i.e.~that each new mutation is either fixed or lost before the next mutation enters
the population. Because the time during which a mutation segregates in such a  population is much shorter than the waiting time between new mutations,
we neglect the time that a mutation segregates and simply model the population as monomorphic, jumping from genotype to genotype
at the birth of each new mutation destined for fixation \citep{Iwasa88,Berg04,Sella05,McCandlish13f}. 

We use the standard model for a population evolving under weak mutation in continuous time. More formally, we model evolution as a continuous time 
Markov chain with rate matrix $\mathbf{Q}$, where
\begin{equation}
\mathbf{Q}(i,j)=\begin{cases}
\frac{\mathbf{F}(j)-\mathbf{F}(i)}{1-e^{-\left(\mathbf{F}(j)-\mathbf{F}(i)\right)}}\mathbf{Q_{M}}(i,j) & \mbox{for } i\neq j \\
-\sum_{k\neq i}\mathbf{Q}(i,k) & \mbox{for } i=j
\label{eq:Qmat}
\end{cases}
\end{equation}
and $\mathbf{Q_{M}}$ is the mutational rate matrix. We assume that forward mutations arise at site $l$ as a Poisson process with rate $\mu_{l}$ and back mutations arise as at site $l$ as a Poisson process with rate $\nu_{l}$. Thus, $\mathbf{Q_{M}}(i,j)$ for $i\neq j$ is equal to $\mu_{l}>0$ if genotype $j$ can arise from 
genotype $i$ by a forward mutation at site $l$, $\nu_{l}>0$ if genotype $j$ can arise from genotype $i$ by a back mutation at site $l$, 
and $0$ otherwise; the $\mathbf{Q_{M}}(i,i)$ are chosen so that the row sums are zero. While for convenience the above expression is based on the classical 
approximation to the probability of fixation of a new mutation in the diffusion limit~\citep{Fisher30,Wright31}, our results can easily be extended to hold exactly
in the limit of weak mutation for a population of finite size $N$ evolving under a Moran process by using the appropriate exact expression for
the probability of fixation~\citep{Moran59}.

We define a non-epistatic fitness landscape~\citep{Phillips08,deVisser11,Szendro13,Weinreich13,deVisser14} as one in which each site makes an additive contribution to fitness (recall that we are
working in Malthusian fitness, so that this corresponds to a multiplicative landscape in Wrightean fitness). More formally, for a non-epistatic fitness
landscape each site $l$ is associated with a value $S_{l}$ such that, for any ordered pair of genotypes $i$ and $j$ differing by a forward
mutation at site $l$, we have $\mathbf{F}(j)-\mathbf{F}(i)=S_{l}$. Under this definition, the non-epistatic fitness landscapes are precisely those landscapes for
which sites evolve independently of each other. This is because the forward substitution rate at site $l$ is always 
\begin{equation}
\alpha_{l}=\frac{S_{l}}{1-e^{-S_{l}}}\,\mu_{l}
\end{equation}
and the corresponding backwards substitution rate is
\begin{equation}
\beta_{l}=\frac{-S_{l}}{1-e^{-(-S_{l})}}\,\nu_{l}.
\end{equation}
Thus, for a non-epistatic fitness landscape the evolutionary dynamics at site $l$ depend only the state of site $l$ and not on the states of the other sites.

\subsection{Mean fitness trajectories}

\subsubsection{Formal results}

\begin{figure} \center \includegraphics[width=9cm]{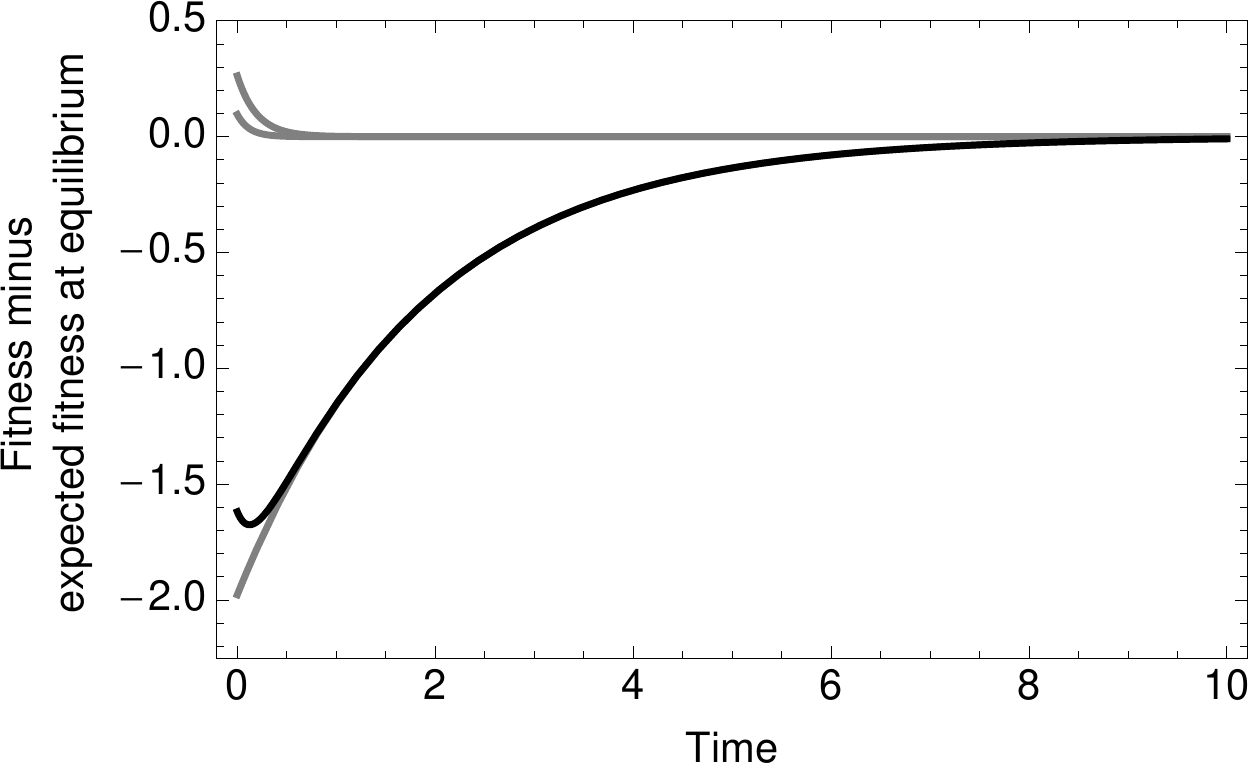}
\caption{
The mean fitness trajectory shared by the two fitness landscapes in Figure~\ref{fig:illustration} (black) together with each of the corresponding 
exponentially decaying deviations from the equilibrium expected fitness (gray) that can be combined together to compose the trajectory. 
There are two positive deviations from the equilibrium expected 
fitness that decay very rapidly, producing the small dip in fitness at short times, and a much more slowly decaying negative deviation from the equilibrium distribution 
that corresponds to crossing the fitness valley in the epistatic landscape or to having a substitution at the third site in the non-epistatic
landscape.
}
\label{fig:decomp} \end{figure}

Suppose a population is initially fixed for some genotype $i$, with fitness $F(i)$, at time $0$. At any time $t$ in the future, the population has some 
probability of being fixed for each other genotype $j$, with fitness $F(j)$. We can therefore ask: What is the expected fitness of the population at time $t$? 
The course of the expected fitness over time is called the mean fitness trajectory, which we write as $f(t)$. In the Supporting Information, we show
that the mean fitness trajectory can always be written in the form:
\begin{equation}
\label{eq:fitnessdecomp}
f(t)=f(\infty)+\sum_{k=2}^{n} c_{k}e^{-\lambda_{k} t}
\end{equation}
for some constants $c_{2},\ldots,c_{n}$, with $\lambda_{2},\ldots,\lambda_{n}>0$ and $n$ denoting the number of genotypes in the fitness landscape.
In other words, for an arbitrary fitness landscape, including all epistatic landscapes, the mean fitness trajectory can always be expressed as a sum of exponentially decaying 
deviations from the equilibrium mean fitness, $f(\infty)$. Figure~\ref{fig:decomp} illustrates this decomposition for the fitness trajectory 
shared by the two fitness landscapes in Figure~\ref{fig:illustration}A.

Now, let us restrict our attention to non-epistatic landscapes, and consider what types of fitness trajectories can arise. Because fitness is additive over sites in such a landscape, we can write the fitness trajectory as a sum over sites. In particular, using the standard solution for a two-state Markov chain, the fitness trajectory for a non-epistatic landscape is given by:
\begin{equation}
\label{eq:nonepfit}
f(t) = f(0)+\sum_{l} S_{l}\,\frac{\alpha_{l}}{\alpha_{l}+\beta_{l}}\left(1-e^{-(\alpha_{l}+\beta_{l})t}\right)
\end{equation}
where $f(0)$ is the initial fitness and we assume (without loss of generality) that the population begins fixed for the first of the two states for each site at $t=0$. 
Rewriting 
Equation~\ref{eq:nonepfit} in the same form as Equation~\ref{eq:fitnessdecomp}, we have:
\begin{equation}
\label{eq:nonepdecomp}
f(t)=\left( f(0)+\sum_{l} S_{l}\,\frac{\alpha_{l}}{\alpha_{l}+\beta_{l}}\right)+\sum_{l} -S_{l}  \frac{\alpha_{l}}{\alpha_{l}+\beta_{l}}\,e^{-(\alpha_{l}+\beta_{l})t}.
\end{equation}

We now arrive at our main result. Comparing Equations~\ref{eq:fitnessdecomp} and~\ref{eq:nonepdecomp}, we see that given an arbitrary epistatic fitness 
landscape we can always construct a non-epistatic fitness landscape that will produce a fitness trajectory of the same shape,
provided we can choose values $S_{l}, \mu_{l}$ and $\nu_{l}$ for each $k$ such that
\begin{equation}
\label{eq:sat1}
c_{k}=-S_{l}  \frac{\alpha_{l}}{\alpha_{l}+\beta_{l}}
\end{equation}
and
\begin{equation}
\label{eq:sat2}
\lambda_{k}=\alpha_{l}+\beta_{l}.
\end{equation}
Indeed, such choices can always be made. For instance, one solution is $\alpha_{l}=\beta_{l}=\lambda_{k}/2$, $S_{l}=-2c_{k}$ 
and $\mu_{l}=\alpha_{l}\left(S_{l}/(1-e^{-S_{l}})\right)^{-1}$, $\nu_{l}=\beta_{l}\left(-S_{l}/(1-e^{-(-S_{l})})\right)^{-1}$, but there are an infinite number of such solutions.

Thus far we have shown that given a fitness trajectory from an epistatic fitness landscape, we can construct a non-epistatic fitness landscape whose
fitness trajectory has the same shape, i.e.~one that differs from the target fitness trajectory by an additive constant. In order to match the fitness
trajectory exactly, we need to be able to choose the term in parentheses in Equation~\ref{eq:nonepdecomp} to be equal to $f(\infty)$. 
But this, too, is always possible to do, because we can freely
choose the initial fitness $f(0)$ and the sum over sites has some definite value fixed by our previous choice of $S_{l}, \mu_{l}$ and $\nu_{l}$, i.e.~we can always 
use the solution
\begin{equation}
f(0)=f(\infty)-\sum_{l} S_{l}\,\frac{\alpha_{l}}{\alpha_{l}+\beta_{l}}.
\end{equation}

To summarize, we have shown that given an arbitrary epistatic fitness landscape
and choice of initial genotype, we can always construct a non-epistatic fitness
landscape that will produce the exact same mean fitness trajectory. In other
words, the presence of epistasis does not expand the range of possible dynamics of
expected fitness gains over time. As a result, one cannot infer whether epistasis
is present or absent from the mean fitness trajectory alone.

\subsubsection{Practical analysis}

The preceding formal analysis suggests a number of natural questions concerning the features of non-epistatic fitness
landscapes that produce the same fitness trajectory as a focal epistatic
landscape. 

First, it is worth noting that that the non-epistatic landscapes constructed above will typically be much larger than the focal epistatic landscape. For instance, if the focal landscape has $L$ sites and $2^{L}$ genotypes, the
corresponding non-epistatic fitness landscape with the same fitness trajectory
produced by the method described above will generically have $2^{L}-1$ sites and therefore
$2^{2^{L}-1}$ genotypes. It is thus natural to ask how closely a non-epistatic landscape of the {\em same} size as the original landscape can match the mean fitness trajectory.

To address this question we considered the class of $LK$
landscapes~\citep{Kauffman89,Kauffman93} with $L=10$. For such landscapes, $L$
controls the number of sites and $K$ controls the ruggedness of the landscape, so
that $K=0$ produces a non-epistatic landscape and $K=L-1$ produces an uncorrelated
(i.e.~house of cards) landscape. We constructed these landscapes as described in
the original publications~\citep{Kauffman89,Kauffman93}, but multiplied all
fitnesses by a constant chosen so that for the non-epistatic case ($K=0$) the
expected value of $|S_{l}|$ equals 5; and we set the forward- and back-mutation rates
at each site equal to unity. For each landscape we constructed, we picked an initial
genotype and then calculated the resulting values for the mean fitness trajectory
at intervals of 0.01 time units, until reaching the final time of 10 units. 
So as to focus on cases of adaptive
evolution, we retained only those landscapes for which the increase in expected fitness
over this time period was at least 10 in units of scaled fitness, and we continued to generate landscapes until
we had 100 such landscapes for each value of $K=1,2$ and $9$. These mean fitness trajectories
were then fit to a model of the form given in Equation~\ref{eq:nonepdecomp} using
non-linear least squares.

\begin{table}
\footnotesize
\begin{center}
\begin{tabular}{ccccc}
$K$  & Mean $R^{2}$ & Mean max error & Mean max \% error & Mean fitness range  \\ \hline
1  & .999995 & .027 & .10 \% & 33.0 \\
2 & .999993 & .029 & .08 \% & 35.5 \\
9 & .999943 & .098 & .38 \% & 30.5 \\
\end{tabular}
\end{center}
\caption{\label{tab:unconst}\textbf{Fitting the mean fitness trajectories from $LK$ landscapes 
using non-epistatic landscapes of the same size.}  The maximal error is the maximum absolute value 
of the error at any time point. The fitness range is the maximum value of the 
mean fitness trajectory at any time point minus the minimum value of the mean fitness trajectory 
at any time point. The percent error is the maximal error divided by the fitness range. }
\end{table}

Our mathematical results above guarantee that we can exactly match the mean fitness
trajectory of 
any such $LK$ landscape using a non-epistatic landscape on $1,023$ sites. 
But how well can we match the mean fitness trajectory using only $10$
sites? Table~\ref{tab:unconst} shows that we can often match the mean fitness
trajectory quite well, even constrained to additive landscapes on the same number
($L=10$) of sites. For
instance, for uncorrelated fitness landscapes ($K=9$), the average $R^{2}$ for our
hundred fits was $.999943$, and the average for the maximum absolute error at any
time point was $.098$ in units of scaled fitness. To put this latter number in
perspective, the mean of the total range of scaled fitnesses displayed during
these mean trajectories was $30.5$, so that the error is only a very small
fraction of the total change in fitness.

In addition to these simulation results, we prove in the Supporting Information that 
given $m$ sites, one can always construct a non-epistatic landscape such that the maximum 
absolute error in the mean fitness trajectory compared to an arbitrary epistatic landscape 
is bounded from above by
\begin{equation}
\frac{1}{m+1}\sqrt{\frac{\operatorname{Var}_{\boldsymbol{\pi}}\mathbf{F}}{\boldsymbol{\pi}(i)}},
\end{equation}
where $\operatorname{Var}_{\boldsymbol{\pi}}\mathbf{F}$ is the variance in fitness at equilibrium 
and $\boldsymbol{\pi}(i)$ is the equilibrium frequency of the initial genotype for the epistatic landscape. 
Furthermore, it is worth noting that many theoretical fitness landscapes have symmetries that 
reduce the number of sites needed to exactly match the mean fitness trajectory using a non-epistatic 
landscape, e.g.~models where the fitnesses and mutation rates depend only on the Hamming distance 
from some focal genotype~\citep{Kimura66,Kondrashov88}, whose mean fitness trajectories can always 
be matched by a non-epistatic landscape of the same size.

Second, one might be concerned that although the method presented above of
constructing a non-epistatic landscape with a specified mean fitness trajectory is
formally valid, the necessary mutation rates and selection coefficients would not
be biologically realistic. To show that one can closely approximate many fitness
trajectories with a small number of site, each of which has a realistic selection
coefficient and mutation rate, we fit the same set of $LK$ landscapes with the
constraint that for each site the forward and backward mutation rates were equal.
For negative terms in the sum of exponentials fit, this constraint uniquely
specifies the corresponding selection coefficient and mutation rate. For positive
terms, this constraint imposes an upper bound on the size of the terms and
otherwise allows two solutions for the pair $\mu_{l}$, $S_{l}$; the pair was
chosen so as to minimize $\mu_{l}$. 

Table~\ref{tab:const} shows that these constrained fits still typically match the
fitness trajectory very closely, if not quite as well as the unconstrained fits in
Table~\ref{tab:const}, e.g. for $K=9$, the mean $R^{2}$ was $.999883$ instead of
$.999943$. Furthermore, the selection coefficients of these fitted landscapes are
reasonable, with mean $|S_{l}|$ approximately $4$ for the
non-epistatic models. The total mutation rates necessary to achieve these fits
were on average approximately 3 times the total mutation rate of the corresponding
epistatic landscapes. 

\begin{sidewaystable}
\footnotesize
\begin{center}
\begin{tabular}{cccccccc}
$K$  & Mean $R^{2}$ & Mean max error & Mean max \% error & Mean fitness range & Mean $|S_{l}|$ & Mean total mutation rate ratio & Mean$[($Max $\mu_{l}$, Min $\mu_{l}$$)$]\\ \hline
1  & .999950 & .101 & .32 \% & 33.0 & 4.1 & 3.1 & (10.8, 0.14) \\
2 & .999942 & .085 & .26 \% & 35.5 & 4.3 & 3.4  & (13.9, 0.06)\\
9 & .999883 & .155 & .69 \% & 30.5 & 4.5 & 3.0 & (11.6, 0.02)\\
\end{tabular}
\end{center}
\caption{\label{tab:const}\textbf{Fitting the mean fitness trajectories from $LK$ landscapes using non-epistatic landscapes of the same size with symmetric mutation rates ($\mu_{l}=\nu_{l}$ for all $l$).}  Maximal error, fitness range and percent error are defined the same way as in Table~\ref{tab:unconst}. The total mutation rate ratio is the total mutation rate in the inferred non-epistatic landscape ($\sum_{l} \mu_{l}$) divided by the total mutation rate in the original epistatic landscape.}
\end{sidewaystable}

Third, there is a question about the role of mutation in the above theory.
While allowing the site-specific mutation rates to be asymmetric is not necessary
to ensure the existence of a non-epistatic landscape that can ``spoof'' the mean
fitness trajectory, it is necessary that the mutation rates are
allowed to differ from site to site. At a mathematical level, this is necessary to
allow the initial deviation from the equilibrium fitness contribution of a site
and the rate at which that site approaches equilibrium to be varied independently.
At an intuitive level, small mutation rates are being used to mimic the effects on
expected fitness of complex aspects epistatic evolutionary dynamics, such as
the waiting time to cross fitness valleys. 

It should be noted that requiring symmetric mutation rates limits the size of the
$c_{k}$ that can be accounted for by a single non-epistatic site (in particular,
the greatest value $c_{k}$ that can be accounted for by a single site is $.278$,
which occurs when $S_{l}=-1.28$; negative values of $c_{k}$ can be matched
regardless of their magnitude). However, one can still construct a non-epistatic
landscape to match the mean fitness trajectory of an arbitrary epistatic landscape
by having multiple sites corresponding to a single term in
Equation~\ref{eq:fitnessdecomp}. In particular, one can generalize
Equation~\ref{eq:sat1} to \begin{equation} c_{k}=-\sum_{l\in \mathcal{L}_{k}}
S_{l}  \frac{\alpha_{l}}{\alpha_{l}+\beta_{l}} \end{equation} where
$\mathcal{L}_{k}$ is the set of sites in the non-epistatic fitness landscape
corresponding to the term $c_{k}e^{-\lambda_{k}t}$ in
Equation~\ref{eq:fitnessdecomp}, and $\alpha_{l}+\beta_{l}=\lambda_{k}$ for all $l
\in \mathcal{L}_{k}$.  Note that this flexibility of using additional sites means
that one can alter the higher moments of the time-dependent fitness distribution
of a non-epistatic fitness landscape while keeping the mean (i.e.~mean fitness
trajectory) unchanged.

\subsection{Variance trajectories}

We have seen that any mean fitness trajectory produced by an epistatic fitness landscape can also be produced by a non-epistatic fitness
landscape. However, the mean fitness trajectory captures only the central tendency of population fitness through time. If we initiate many replicate populations
fixed for the same genotype, then it is likely that some populations will adapt more quickly than others, 
so that there will typically be variation in fitness across
populations at any given time $t>0$. Aside from the mean, discussed above, it is natural to ask whether the presence of epistasis increases the diversity of the possible dynamics of the inter-population variation in fitness.

To make this idea more precise, let us consider the fitness of a population at time $t$ as a random variable. 
The variance of this random variable viewed as a function of time is called the ``variance trajectory'', $v(t)$. 
In other words, the variance trajectory is the time evolution of the second central moment of the fitness distribution, 
across an ensemble of replicate populations.

We would like to know whether  the set of variance trajectories that can be achieved by epistatic fitness landscapes is more diverse than the 
set that can be achieved by non-epistatic fitness landscapes. 
To answer this question, we will first use the standard solution for a two-state Markov chain to write down the variance trajectory for a single site:
\begin{equation}
\label{eq:var}
v(t)=\left(S\frac{\alpha}{\alpha+\beta}\right)^{2}\left( \frac{\beta}{\alpha}\,\left(1-e^{-(\alpha+\beta)t} \right)+\left(e^{-(\alpha+\beta)t}-e^{-2(\alpha+\beta) t}\right)\right),
\end{equation}
assuming without loss of generality that the population starts in the first state. 
It is easy to show that the first derivative of $v(t)$ with respect to time is maximized at $t=0$, 
since the derivatives of both $1-e^{-t}$ and $e^{-t}-e^{-2t}$ are maximized at $t=0$. 
Thus, the rate that variance in fitness increases takes its maximum at $t=0$, which makes sense, because at $t=0$ the 
fitness of the alternative state is maximally different from the current mean fitness, and the increase in the frequency of the 
alternative state is also maximized (because back-substitutions cannot occur at time $t=0$).

The variance trajectory of a non-epistatic fitness landscape is simply the sum of the variance trajectories across all sites (because
variances can be summed when random variable are independent). 
Now, because the slope of the variance trajectory is maximized for each site at $t=0$, 
it follows 
that the slope of the variance trajectory is maximized at $t=0$ for any finite, non-epistatic fitness landscape. 
Furthermore, because the slope is maximized at $t=0$, it follows that the second derivative of the fitness trajectory must be negative at $t=0$. 
In other words, all non-epistatic fitness landscapes share a fundamental qualitative feature: 
their variance trajectories are concave at short times.

Are variance trajectories for all epistatic fitness landscapes also concave at short times? The answer is no. 
For instance, consider a two-site fitness landscape with genotypes ab, Ab, aB, and AB, and $\mu_{l}=\nu_{l}=1$ for both sites, 
with a population initially fixed for genotype ab. Suppose the fitnesses of ab, Ab and aB are all equal but genotype AB has fitness advantage $S$ 
over the other three genotypes. The first derivative of the resulting variance trajectory at $t=0$ (i.e., $v'(0)$) is zero, 
and the second derivative at $t=0$ (i.e. $v''(0)$) is $S^{3}/\left(1-e^{-S}\right)$, which is positive for $S\neq0$. 
Thus, for any such landscape with $S\neq0$, the variance trajectory is convex at short times---a feature that cannot be achieved
by any non-epistatic landscape.

\begin{figure}[t] \center \includegraphics[width=15cm]{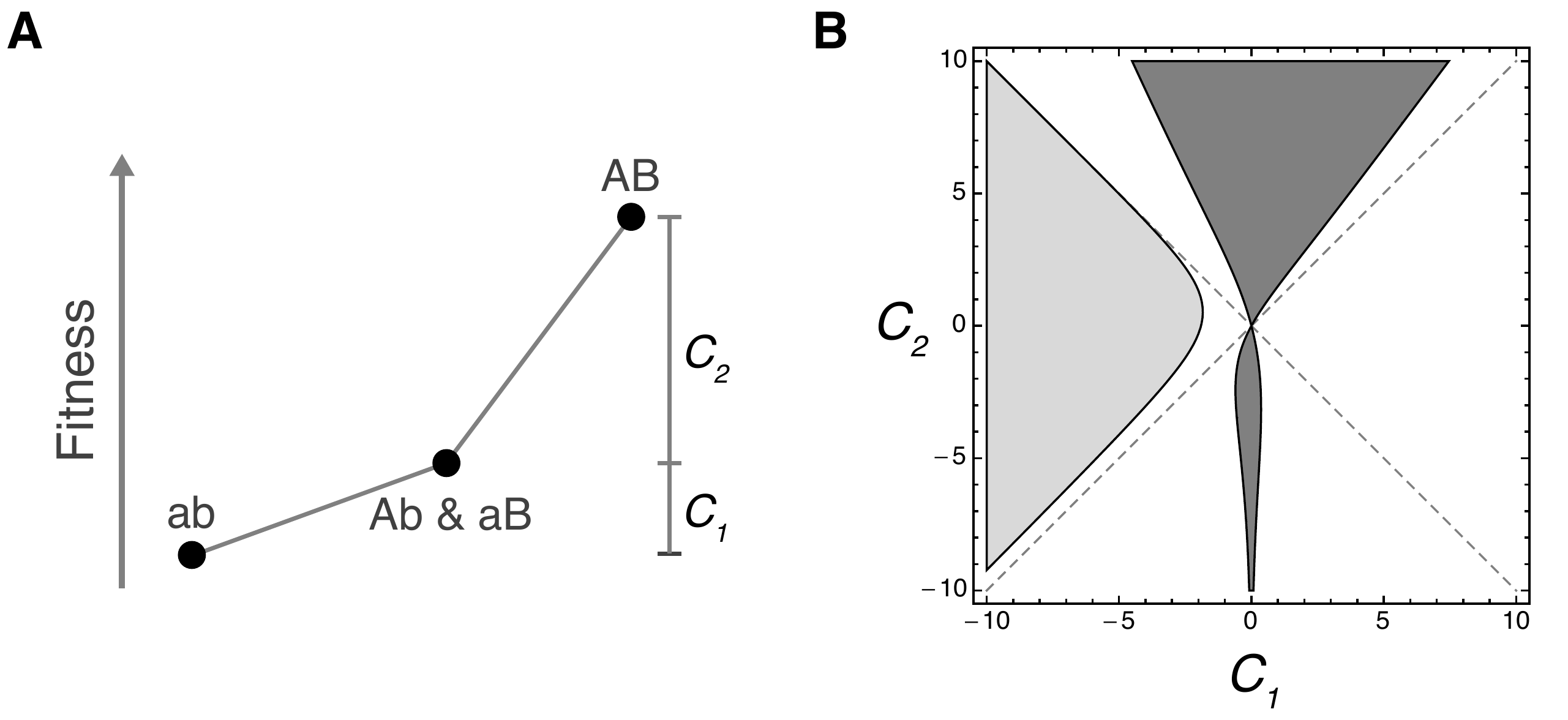}
\caption{
Epistatic fitness landscapes can produce dynamics that differ from all non-epistatic landscapes. (A) We consider two-site landscapes with
genotypes ab, Ab, aB, and AB, with the population initially fixed for ab.
Assigning Ab and aB equal fitnesses, we let $C_{1}$ be the selection coefficient of Ab and aB relative to ab and let $C_{2}$ be the selection coefficient of AB relative to Ab and aB. We set the mutation rates to be $\mu_{l}=\nu_{l}=1$ for both sites.
(B) Properties of two-site fitness landscapes using the parameterization described in the previous panel.  The dark gray region shows the set of landscapes whose variance trajectories are convex at $t=0$. The light gray region shows the set of
landscapes whose equilibrium substitution rates are greater than twice their initial substitution rates. 
The diagonal dashed line with positive slope
shows the set of non-epistatic fitness landscapes, whereas the diagonal dashed line with negative slope shows the set of 
epistatic landscapes whose fitness dynamics cannot be distinguished from the non-epistatic case.
}
\label{fig:2site} \end{figure}

As illustrated by the example above, the range of possible dynamics for the variance trajectory 
is larger for epistatic landscapes than for non-epistatic landscapes. And so it may be possible to infer epistasis from
the pattern of variance in fitness across populations, even though it is impossible to do so from the pattern of mean fitness alone.

To explore the range of epistasic landscapes that produce variance trajectories that are convex at short times, we considered the two-site
landscapes described above, but allowed Ab and aB to have some selection coefficient relative ab, and also allowed AB to have some other selection
coefficient relative to Ab and aB (i.e.~genotypes with equal hamming distances from ab are assigned equal fitnesses). The dark gray region in
Figure~\ref{fig:2site} shows the subset of landscapes whose the variance trajectories are convex at short times. Although this region is primarily
composed of landscapes with positive epistasis (above the line y=x), it is also possible to have convex variance trajectories for landscapes
with negative epistasis (below the line y=x) when the selection coefficient of the first mutation is small.

We have seen that some epistatic fitness landscapes produce variance trajectories that cannot be achieved by any non-epistatic fitness
landscapes. Thus, we have shown that one can sometimes tell that a fitness landscape is epistatic by observing its time-dependent fitness
distribution. However, one might wonder whether given the time-dependent fitness distribution it is always possible to distinguish epistatic from
non-epistatic fitness landscapes. The answer is no. As a counter-example, consider again the two-site case in
which Ab and aB have equal fitness. If the selection coefficient of ab relative to Ab and aB is equal to the selection coefficient of AB relative to
Ab and aB, then the entire fitness distribution can be matched by a single-site fitness landscape.  These landscapes are illustrated by the dashed
line with negative slope in Figure~\ref{fig:2site}. Thus, there is no characteristic of the time-dependent fitness distribution that can be used to
distinguish all epistatic landscapes from all non-epistatic fitness landscapes.

Another natural question, in light of our earlier results on fitness trajectories, concerns the relationship between the mean fitness trajectory and the variance trajectory, for a given landscape.  Within the class of
non-epistatic landscapes, it is easy to show that the variance and mean fitness trajectories can be modified essentially independently of each other. This
is because the variance in fitness can be made arbitrarily small while preserving the fitness trajectory by replacing single sites of large fitness
effect with many sites of small effects; on the other hand, the variance in fitness can be made arbitrarily large without altering the fitness trajectory by
constructing pairs of sites whose site-specific fitness trajectories cancel each other out, but which still contribute to the time-evolution of the
variance. As a result, considering the fitness and variance trajectories jointly is likely to provide little more information about the presence of epistasis than considering the variance trajectory alone.

\subsection{Substitution trajectories}

Changes in fitness during adaptation are the result of substitutions -- that is, mutations at individual sites that eventually reach fixation in the population.
Therefore, aside from studying the expected fitness of a population,
it is also interesting to consider the the number of substitutions that accumulate in a population over time. For
instance, consider the time-evolution of the expected number of substitutions that have accumulated in the population by time $t$, what we call the
``substitution trajectory'', $s(t)$~\citep{Kryazhimskiy09}.  Just as we did for mean fitness trajectories, we want to ask whether the set of possible
substitution trajectories for epistatic fitness landscapes is more diverse than the set of possible substitution trajectories in the absence of
epistasis.

To study the substitution trajectory, it is helpful to note that the derivative of the substitution trajectory is equal to the expected substitution
rate at time $t$, which we will write as $q(t)$. That is, $s'(t)=q(t)$. Because no substitutions have accumulated at time $0$, this relation means
that the substitution trajectory is fully specified by the time-dependent expected rates of substitution. 

Consider the time-dependent rate of substitution at a single site. Assuming without loss of generality that the population begins in the first state, and using the standard solution for a two-state Markov chain,
the expected substitution rate is given by:
\begin{equation}
q(t)=\frac{2\alpha \beta}{\alpha+\beta}-\frac{\alpha}{\alpha+\beta}(\beta-\alpha)e^{-(\alpha+\beta)t},
\end{equation}
where the first term on the right-hand side is the equlibrium rate of substitution, $q(\infty)$. 
The initial substitution rate is $\alpha$, and so the ratio between the equilibrium substitution rate and the 
initial substitution rate is $q(\infty)/q(0)=2\beta/(\alpha+\beta)$ -- a ratio than can never exceed 2, which is the value achieved in the limit 
as $\beta/\alpha \rightarrow \infty$. Indeed, because the expected substitution rate approaches its equilibrium value monotonically, 
we also have a stronger result: the expected substitution rate can never be less than half of its equilibrium rate, i.e.~$q(\infty)/q(t)\leq 2$.

For a non-epistatic landscape, 
the expected substitution rate is simply a sum of rates at each of its constituent sites.
Using the inequality developed above, we thus have:
\begin{align}
\frac{q(\infty)}{q(t)}&=\frac{ \sum_{l}q_{l}(\infty)}{\sum_{l}q_{l}(t)}\\
 &\leq \frac{ \sum_{l}2\,q_{l}(t)}{\sum_{l}q_{l}(t)}\\
 &=2,
\end{align} 
where $q_{l}(t)$ is the expected substitution rate at the $l$-th site at time $t$. 
In words, for any finite-state non-epistatic fitness landscape, the ratio between the equilibrium rate and the expected rate at any time can never
exceed two. 

For epistatic landscapes, by contrast, it is easy to see that this condition on the substitution rate can be violated.
For instance, consider a fitness landscape with three or more sites in which all genotypes have the same fitness. 
Then, pick an initial genotype, and alter the fitnesses of its mutational neighbors such that these neighbors now have selection coefficient $S$ 
relative to the initial genotypes, where we choose $S$ to be negative. That is, consider a neutral plateau and modify it by constructing a fitness valley around 
the initial genotype. As the depth of this valley increases (i.e.~as $S$ approaches $-\infty$), the initial substitution rate converges to 0, 
while the equilibrium substitution rate approaches some non-zero constant. This means that by choosing $S$ to be sufficiently large and negative
the ratio between the equilibrium substitution rate and the initial substitution rate can be made arbitrarily large and, in particular, larger than two. 
(While no adaptation occurs in this example -- the mean equilibrium fitness is lower than the initial fitness -- this defect is easy to correct by 
giving a fitness advantage to genotypes of distance two or more from the initial genotype). Thus, we conclude 
that the set of possible substitution trajectories is indeed enlarged by the presence of epistasis.

The light gray region in Figure~\ref{fig:2site} illustrates this fact, by indicating the set of two-site fitness landscapes whose ratios of equilibrium to initial 
expected substitution rates exceeds 2.
Roughly speaking, this region corresponds
to landscapes with a fitness valley, with population initialized on the fitter of the two peaks. Note that the light gray region does not extend all the way to
the landscapes in which the two peaks have equal heights: these landscapes (dashed diagonal line with negative slope) have substitution trajectories of
precisely the same form as a single site, and therefore the ratio of rates must be less than or equal to 2 along this line. More generally, the
adaptive situation most likely to produce initial substitution rates that are less than half the equilibrium substitution rate is a population that is
currently at an isolated local fitness maximum, but which, after crossing a fitness valley, will spend most of its time on a broad, high fitness plateau.

Why is the range of dynamics of mean fitness identical for epistatic and non-epistatic landscape, but not the range of dynamics of the mean number
of substitutions? One way to understand these results is to notice that, because mutation can oppose or augment selection, two non-epistatic landscapes
with different fitness functions and different mutation rates might still have the same evolutionary dynamics in genotype space (i.e.~the same rate
matrix, $\mathbf{Q}$). As a consequence, for each non-epistatic matrix $\mathbf{Q}$ and choice of starting genotype, there is a large class of possible fitness
trajectories, determined by the choices of the site-specific selection coefficients $S_{l}$. In contrast, having specified the matrix $\mathbf{Q}$ and the
initial genotype completely determines the substitution trajectory. The extra flexibility produced by choosing $\mathbf{Q}$ and the $S_{l}$
independently allows non-epistatic fitness landscapes to produce fitness trajectories whose dynamics are as general as the time-evolution of
the expectation of an arbitrary function defined on an arbitrary finite-state reversible Markov chain.

For completeness, we can also consider the ensemble variance in substitution rate as a function of time. The derivative of this trajectory is
maximized at $t=0$, just as the derivative of the variance in fitness is as well.  While this criterion can be used to identify some fitness
landscapes as epistatic, the time evolution of the variance in substitution rate is much more difficult to observe than the time evolution of the mean
substitution rate or the time evolution of the fitness distribution, and so we will not discuss the matter further here.

\subsection{Equilibrial dynamics}

Although our main focus has been adaptation, it is also interesting to consider whether epistatic and non-epistatic landscapes differ in the range of
dynamics they can produce at equilibrium, i.e.~in the limit of long times when all influence of the choice of initial genotype has been lost.  We
study the equilibrial dynamics by again considering an ensemble of replicate populations. However, instead of assuming that all of these
populations are initially fixed at a single genotype, we assume that the initial genotype for each population is drawn from the equilibrium
distribution, that is, the distribution that gives the probability of a population being fixed for any given genotype in the limit of long times. 

An ensemble of populations that is initially distributed according to the equilibrium distribution will continue to be described by the
equilibrium distribution at all future times. Hence the expected fitness, the variance for fitness, and indeed all moments of the fitness
distribution are constant in time. Indeed, the equilibrium fitness distribution is determined solely by the fitnesses of the individual genotypes together with
their equilibrium frequencies, and it is therefore independent of the structure of the fitness landscape in the sense that the structure of
mutational adjacency is irrelevant~\citep[see][pg.~1547]{McCandlish11}. Because the equilibrium distribution
remains constant in time and each genotype has its own substitution rate, substitutions likewise accumulate at a constant rate across the ensemble as
a whole.

However, while the fitness distribution across the ensemble remains constant in time, individual populations in the ensemble will still experience changes in fitness. We can study the structure of these changes by studying the correlations between the fitness of a population at one time and its fitness at another. In particular, we consider the covariance in fitness between time $t'$ and some later time $t'+t$, where again the genotype at time $0$ is drawn from the
equilibrium distribution. Viewed as a function of the difference, $t$ between these two times, this covariance is known as the equilibrium
autocovariance for fitness, denoted $a(t)$. In the Supporting Information, we show that the equilibrium autocovariance for an
arbitrary epistatic fitness landscape has the form:
\begin{equation}
\label{eq:autoco}
a(t)=\sum_{k=2}^{n} d_{k} e^{-\lambda_{k}t},
\end{equation}
where $d_{2},d_{3},\dots \geq 0$ and $\lambda_{2},\lambda_{3},\ldots>0$. 

For comparison, let us know consider the autocovariance for non-epistatic landscapes, considering first a landscape with a single site with selection
coefficient $S$. In this case the equilibrium autocovariance for fitness is given by
\begin{equation}
a(t)=S^{2}\frac{\alpha \beta}{(\alpha+\beta)^{2}}e^{-(\alpha+\beta)t}.
\end{equation}
The autocovariance of a sum of independent processes is the sum of the corresponding autocovariances, and the 
term $S^{2}\alpha \beta/(\alpha+\beta)^{2}$ can assume any non-negative value even with a fixed value of $\lambda_{k}=\alpha+\beta$. This implies
that, given the 
equilibrium autocovariance function for an epistatic fitness landscape, one can always construct a non-epistatic landscape with an identical equilibrium autocovariance
function by assigning one site to correspond to each term in Equation~\ref{eq:autoco}. Thus, while the presence of epistasis increases the possible dynamics for the second
moment of fitness for an adapting population, epistatic and non-epistatic fitness landscapes have the same range of possible dynamics for the equilibrium autocovariance in fitness.

\section{Discussion}

How does the structure of the fitness landscape influence the dynamics of adaptation? 
Here we have studied this question by identifying dynamical phenomena that can occur when epistasis
is present but that cannot occur when epistasis is absent. We have considered the evolution
of populations under weak
mutation on arbitrary fitness landscapes defined on a finite number of bi-allelic sites. For each of several basic descriptors of
adaptation---e.g.~the expected fitness or number of substitutions accrued over time---we have asked whether the dynamics that are
possible on epistatic fitness landscapes are more diverse than those possible on non-epistatic landscapes. The results are surprisingly heterogeneous.

The most basic and essential descriptor of adaptation is the mean fitness trajectory---that is, the expected pattern of fitness gains over time. In
contrast to the received wisdom that the presence or specific form of epistasis can alter the mean pattern of adaptation over time
~\cite[e.g.~epistasis ``accelerating" or ``de-accelerating" adaptation,][]{Chou11,Khan11,Kryazhimskiy11b}, we have shown that the set of possible mean fitness trajectories for epistatic fitness landscapes is no more
diverse than for non-epistatic fitness landscapes. In particular, any mean fitness trajectory that can be achieved on an epistatic fitness landscape can
also be acheived by an infinite number of non-epistatic fitness landscapes. Furthermore, while our analytical results show that the number of sites needed to {\em exactly} match the mean fitness trajectory with a non-epistatic fitness landscape is typically much larger than the number of sites in the original fitness landscape, our numerical results on $LK$ landscapes~\citep{Kauffman87,Kauffman89,Kauffman93} suggest that the mean fitness trajectories of even highly rugged fitness landscapes may often be closely approximated by the mean fitness trajectories of non-epistatic landscapes with the same number of sites.

In contrast to the mean fitness trajectory, we have shown that the time-evolution of the
variance in fitness across populations can display qualitatively different behavior on epistatic
landscapes than can be achieved in the absence of epistasis. Likewise, the pattern of the expected number of substitutions
accrued over time can also be qualitatively different on epistatic fitness landscapes than possible on non-epistatic landscapes. 

These results have implications for efforts to infer the prevailing form of epistasis by experimentally observing the evolutionary dynamics in an ensemble of
replicate populations~\citep{Lenski91,Lenski94,Kryazhimskiy09,Wiser13, Good14}. Such an approach is appealing because while epistasis is easy to detect by considering combinations of mutations directly~\citep[e.g.][]{Chou11,Khan11}, inferences about the prevailing form of epistasis based on small samples of mutations can be strongly misleading~\citep{Blanquart14}, particularly when the mutations chosen are those that fixed during adaptation~\citep{Draghi13}. Efforts to infer the global form of epistasis using random mutagenesis~\citep[e.g.][]{Olson14,Bank15} suffer from other difficulties, such as only giving information about epistasis in a local region of sequence space or between mutations that are too deleterious to ever fix. Inferring the form of epistasis from the trajectories observed during experimental evolution overcomes these difficulties because the mutations sampled by evolving populations 
are precisely the mutations most relevant to the evolutionary process.

Existing approaches to infer the form of epistasis from experimental trajectories
typically consist of fitting a handful of simple epistatic and non-epistatic
models to the observed data~\citep[e.g.][]{Kryazhimskiy09,Wiser13, Good14}.
However, as emphasized by~\citet{Frank14}, all we can really infer from such an
approach is the set of models consistent with the observed dynamics. Our results
show that, for populations under weak mutation, the shape of the mean fitness
trajectory alone can never be used to infer the presence or form of epistasis,
even with an unlimited number of replicate experimental populations.  Epistasis is
not identifiable by these type of data because any mean fitness trajectory that
can occur on an epistatic landscape is consistent with some non-epistatic landscape as well.
On the other hand, the variance and substitution trajectories can in principle
sometimes allow us to conclude that epistasis must be present. Indeed, our results
on the variance trajectory confirm the conjecture by~\citet{Lenski91} that the
time-evolution of the variance in fitness across replicate populations can be used
to detect epistasis. The simple intuition underlying our analysis is that positive
epistasis can cause the slope of the variance trajectory to increase as time
elapses, whereas when epistasis is absent the slope of the variance trajectory is
always maximized at $t=0$.

While our results on variance and substitution trajectories suggest possibilities
for empirical tests of epistasis, there still remain several obstacles to
developing rigorous and practical tests for the presence and form of epistasis
from these types of data. First, the results presented here are based on the
assumption that we have access to an infinite ensemble of individual populations,
whereas in practice we only have a finite sample of populations. Such a finite
sample produces a number of technical complications, such as the fact that errors
in the estimation of the expected trajectories will be correlated across time
points~\citep{Good14}. Another complication is that, strictly speaking, under our
model the sample means will be piecewise constant, and therefore the derivatives
that we have studied here cannot be calculated directly from the sample means.
Second, there is the issue of measurement error, which we have neglected in our
treatment. Third, there are questions of power. While Figure~\ref{fig:2site} shows
that many simple fitness landscapes would produce substitution and variance
trajectories incompatible with a non-epistatic model, we suspect our compatibility
criterion for the substitution trajectory may be less informative in more
realistic landscapes, both because of neutral substitutions (which will make the
initial and equilibrium substitution rates more similar) and because of the
general intuition that substitution rates should be decreasing during adaptation
as the supply of beneficial mutations is exhausted. Tests of epistasis based on the variance
trajectory or a combination of the substitution and fitness
trajectories~\citep[c.f.][]{Good14} may provide better avenues for future
research. Indeed, the variance trajectory may be particularly well-suited for detecting situations where neutral potentiating mutations are required before substantial adaptation is possible~\citep{AWagner08b,AWagner11,Draghi10,Bloom10}.
 Finally, while our results here assume that mutation is weak, most
experimental evolution involves large microbial populations in the regime of
clonal interference~\citep{Gerrish98}.  Thus, it is possible that existing
experimental mean fitness trajectories~\citep[e.g.][]{Wiser13} may contain
information about the presence of epistasis. However,  without a comparable
demonstration that epistasis is indeed identifiable from the mean fitness
trajectory for such populations, our negative results under weak mutation suggest
extreme caution in using empirical mean fitness trajectories to argue for or
against the hypothesis that epistasis is present or to estimate the prevailing
form of epistasis.

Two recent theoretical studies have also analyzed the relationship between the
presence or absence of epistasis and the dynamics of adaptation under weak
mutation. \citet{Kryazhimskiy09} considered the space of fitness landscapes in
which the distribution of mutational effects on fitness (DFE) is solely a function
of the current fitness of the population, and they concluded that it is possible
to identify epistasis from the mean fitness trajectory. On the whole, the class of
models studied by~\citet{Kryazhimskiy09} is much broader than the one considered
here (since any finite-state fitness landscape can be arbitrarily well
approximated in their framework so long as each genotype has a unique fitness),
and it includes many models that are inconsistent with finite-site landscapes. In
particular, ~\citet{Kryazhimskiy09} considered a fitness landscape to be
non-epistatic if its DFE is independent of the population's fitness. Such a
situation can never arise on a non-trivial finite-site landscape, because the DFE
must be entirely negative at the fittest genotype and entirely positive at the
least-fit genotype. Our results are thus complementary to those of
\citet{Kryazhimskiy09}, and we conclude that while the shape of the fitness
trajectory may be informative in distinguishing between various models in their
broader class, it is not informative for the narrower set of models corresponding
to finite-site fitness landscapes.  It is also worth noting that the analytical
results presented by \citet{Kryazhimskiy09} are approximations that hold only for
a relatively limited subset of models within this broader
class~\citep[see][pp.~124--127]{VanKampen07}, whereas the analytical results
presented here are exact and apply to arbitrary fitness landscapes with a finite
number of biallelic sites.

\citet{Good14} also present results on the diversity of mean fitness and substitution trajectories that can be produced by non-epistatic fitness landscapes, but they assume that selection is strong so that deleterious mutations cannot fix. Under strong selection and weak mutation \citet{Good14} found that non-epistatic landscapes can produce only a subset of possible mean fitness trajectories (these trajectories correspond to the case with all $c_{k}\leq 0$ in our Equation~\ref{eq:fitnessdecomp}). In contrast, here we have allowed for the possibility of deleterious substitutions and we have shown that any mean fitness trajectory that can be produced by an epistatic landscape can also be produced by a non-epistatic landscape.

This difference arises from the assumption of strong selection, which changes the basic character of the evolutionary dynamics~\citep[see also][]{McCandlish13e}. Under our model, each site in a non-epistatic fitness landscape independently approaches a mutation-selection-drift equilibrium~\citep{Bulmer91}. At this equilibrium, the expected fitness and substitution rate can be either lower or higher than the initial fitness and substitution rate. In contrast, under strong-selection weak-mutation, each site in a non-epistatic fitness landscape results in exactly one beneficial substitution, and all evolution comes to a halt when the population reaches the fittest genotype. Thus, the rate of adaptation must be decreasing in time as the finite supply of beneficial mutations becomes exhausted; the same is true of the substitution rate. This greatly constrains the set of mean fitness and substitution trajectories that can be produced by non-epistatic fitness landscapes under strong-selection weak-mutation. \citet{Good14}~also derive a necessary relation between the substitution and mean fitness trajectories for non-epistatic landscapes when mutation rates are uniform across sites, but show in their supplementary material that this relationship breaks down if different sites are permitted to have different mutation rates.

While most of our results have focused on adaptive evolution, we also studied the nearly-neutral dynamics of a population
evolving at equilibrium on a time-invariant fitness landscape. In particular, we showed that the autocovariance function for fitness of
such a population cannot be used to determine whether a fitness landscape is epistatic or not.  This result is in contrast to the autocovariance
function of a completely random walk on the space of genotypes, whose characteristics have long been used to quantify the ``ruggedness'' of fitness
landscapes~\citep{Weinberger90,Weinberger91,Stadler03}. 

One potential limitation of our analysis is that we have considered fitness
landscapes composed of only bi-allelic sites.  This assumption does not, in fact,
influence our results on the space of dynamics possible under epistatic
landscapes. This is because, as shown in the Supporting Information, our
results for epistatic fitness landscapes hold for any time-independent,
finite-state fitness landscape  whose neutral mutational dynamics take the form of
a reversible Markov chain~\citep[see, e.g.][]{Sella05,McCandlish11}.
Thus, our results on epistatic dynamics apply also to models with more than two
alleles per site (so long as the mutational dynamics within a site form a
reversible Markov chain); and, indeed, they even apply when the genotypic space
cannot be decomposed into individual sites. But our assumption of bi-allelic sites
does influence our analysis of non-epistatic models, because our strategy for
determining the behavior of such models has been to sum over the dynamics of
independently evolving sites. The dynamics at a single site can be more complex
when there are more than two alleles, and so the dynamics that are possible under
multi-allelic finite-site non-epistatic fitness landscapes are more diverse than
those described here for non-epistatic models with bi-allelic sites.  Thus, all
our negative results concerning whether epistatic models have more diverse
dynamics than non-epistatic models (such as our main result on the mean fitness
trajectory) continue to hold for multi-allelic models, but our positive results
(such as our results on the variance trajectory) may no longer apply.

Another important limitation of our analysis is that we have considered the evolutionary dynamics only under weak mutation. This is because under weak
mutation sites that do not interact epistatically evolve independently. However, for finite polymorphic populations with linked sites, even sites that do not
interact epistatically have dynamics that are non-independent due to hitch-hiking and background selection. This non-independence makes it extremely
difficult to provide a full treatment of even non-epistatic dynamics for finite, polymorphic populations.  In the absence of analytical results, it is
tempting to try to address the role of epistasis for finite polymorphic populations through simulation.  However, the enormity of the space of
possible fitness landscapes means that any such approach will be restricted to a tiny subset of fitness landscapes, and so it cannot definitively
answer the types questions that we have addressed here.

\section{Acknowledgements}

We thank Richard Lenski, Noah Ribeck, Bj{\o}rn {\O}stman, Benjamin Good and Michael Desai for helpful
discussion, and Thomas Lenormand, Sylvain Gl\'emin and two anonymous reviewers for their comments on the manuscript. We acknowledge funding from the Burroughs Wellcome Fund,
the David and Lucile Packard Foundation, the James S.~McDonnell
Foundation, the Alfred P.~Sloan Foundation, the U.S.~Department of the
Interior (D12AP00025), and the U.S.~Army Research Office (W911NF-12-1-0552).

\pagebreak

\bibliographystyle{Evolution} \bibliography{MainBibtexDatabase}

\pagebreak

%\appendix \def\thesection{Appendix% \arabic{section}
%}
%\renewcommand{\theequation}{A\arabic{equation}} \setcounter{equation}{0}

\clearpage \pagebreak

\setcounter{page}{1} \setcounter{equation}{0} \setcounter{theorem}{0}

\appendix \def\thesection{Supporting Information S\arabic{section}}
\renewcommand{\theequation}{S\arabic{equation}}
\renewcommand{\thetheorem}{S\arabic{theorem}}

\section*{Supporting Information} 

\label{sec:eigendecomposition}

We proceed in somewhat more generality than in the main text. Suppose that evolution under mutation alone proceeds as a reversible, continuous-time Markov chain on a finite state space with rate matrix (infinitesimal generator) $\mathbf{Q_{M}}$ and equilibrium distribution $\boldsymbol{\pi_\mathbf{M}}$. If the scaled Malthusian fitness of genotype $i$ is given by $\mathbf{F}(i)$, then evolution under weak mutation is a Markov chain with rate matrix $\mathbf{Q}$ whose $i,j$-th entry is:
\begin{equation}
\mathbf{Q}(i,j)=\begin{cases}
\frac{\mathbf{F}(j)-\mathbf{F}(i)}{1-e^{-\left(\mathbf{F}(j)-\mathbf{F}(i)\right)}}\mathbf{Q_{M}}(i,j) & \mbox{for } i\neq j \\
-\sum_{k\neq i}\mathbf{Q}(i,k) & \mbox{for } i=j.
\end{cases}
\end{equation}
It is easy to verify that the equilibrium distribution of this chain is given by the vector $\boldsymbol{\pi}$, where $\boldsymbol{\pi}(i) \propto   \boldsymbol{\pi_\mathbf{M}}(i) \,e^{\mathbf{F}(i)}$, and that this equilibrium satisfies detailed balance, so that the chain defined by $\mathbf{Q}$ is also reversible. Note that the more limited definition of $\mathbf{Q_{M}}$ in the main text based on some finite number of bi-allelic sites with non-zero forward and reverse mutation rates necessarily results in a reversible Markov chain, since it is simply the rate matrix for a collection of independent two-state chains with non-zero transition rates, and any two-state continuous-time chain with non-zero transition rates is reversible.

Because the Markov chain defined by $\mathbf{Q}$ is reversible, the definition of detailed balance implies that the matrix $\mathbf{D}_{\boldsymbol{\pi}}^{1/2}\mathbf{Q}\mathbf{D}_{\boldsymbol{\pi}}^{-1/2}$ is symmetric, where $\mathbf{D_{x}}$ is the diagonal matrix whose diagonal entries are given by the vector $\mathbf{x}$. We can thus expand $\mathbf{D}_{\boldsymbol{\pi}}^{1/2}\mathbf{Q}\mathbf{D}_{\boldsymbol{\pi}}^{-1/2}$ in terms of its eigenvalues and eigenvectors as
\begin{equation}
-\mathbf{D}_{\boldsymbol{\pi}}^{1/2}\mathbf{Q}\mathbf{D}_{\boldsymbol{\pi}}^{-1/2} = \sum_{k=1}^{n}\lambda_{k}\mathbf{u}_{k} \transpose{\mathbf{u}_{k}},
\end{equation}
where $0=\lambda_{1}<\lambda_{2}\leq \lambda_{3}\leq \ldots \leq \lambda_{n}$ are the eigenvalues of $-\mathbf{D}_{\boldsymbol{\pi}}^{1/2}\mathbf{Q}\mathbf{D}_{\boldsymbol{\pi}}^{-1/2} $ and the eigenvectors $\mathbf{u}_{k}$ form an orthonormal basis of $\mathbb{R}^{n}$.  Multiplying the above equation by $\mathbf{D}_{\boldsymbol{\pi}}^{-1/2}$ from the left and $\mathbf{D}_{\boldsymbol{\pi}}^{1/2}$ from the right, then gives us: 
\begin{equation}
-\mathbf{Q}=\sum_{k=1}^{n} \lambda_{k}\, \mathbf{r}_{k} \, \transpose{\mathbf{l}_{k}},
\end{equation}
where $\mathbf{l}_{k}=\mathbf{D}_{\boldsymbol{\pi}}^{1/2} \mathbf{u}_{k}$ and $\mathbf{r}_{k}=\mathbf{D}_{\boldsymbol{\pi}}^{-1/2} \mathbf{u}_{k}$ are the left and right eigenvectors of $-\mathbf{Q}$ associated with $\lambda_{k}$. Note that $\transpose{\mathbf{l}_{k}} \mathbf{D}_{\boldsymbol{\pi}}^{-1} \mathbf{l}_{m}=\transpose{\mathbf{r}_{k}} \mathbf{D}_{\boldsymbol{\pi}} \mathbf{r}_{m}=1$ for $k=m$ and $0$ otherwise, and $\mathbf{l}_{k}(i)=\boldsymbol{\pi}(i)\mathbf{r}_{k}(i)$.

The transition probabilities for the Markov chain can then be written in terms of this expansion of $\mathbf{Q}$. In particular, let $\mathbf{P}_{t}$ be the matrix whose $i,j$-th element is the probability that a population that begins at time $0$ fixed for genotype $i$ is fixed for genotype $j$ at time $t$. Then we can write:
\begin{equation}
\label{eq:eigendecomp}
\mathbf{P}_{t}(i,j)=\sum_{k=1}^{n}e^{-\lambda_{k}t} \mathbf{r}_{k} (i) \mathbf{l}_{k} (j).
\end{equation}
As a result, for any function on the state space of the Markov chain, the expected value of that function at time $t$ for a population that begins fixed for genotype $i$ at time $0$ is given by
\begin{equation}
\sum_{k=1}^{n}e^{-\lambda_{k}t} \mathbf{r}_{k} (i)\, \transpose{\mathbf{l}_{k}} \mathbf{g}.
\end{equation}
where $\mathbf{g}(i)$ is the value of the function at genotype $i$. Equation~\ref{eq:fitnessdecomp} follows by choosing $\mathbf{g}(i)=\mathbf{F}(i)$ and noting that because the rows of $\mathbf{Q}$ sum to zero, we must have $\mathbf{r}_{1}=\mathbf{1}$, where $\mathbf{1}$ is the vector of all 1s, for all $i$ and thus $\mathbf{l}_{1}=\boldsymbol{\pi}$.

Next, we turn to deriving the bound on the error of the approximation of a mean fitness trajectory from an arbitrary epistatic fitness landscape using a fitness landscape with $m$ sites. In particular, we will show that for any mean fitness trajectory $f(t)$ produced by an arbitrary, finite fitness landscape whose mutational dynamics take the form of a reversible Markov chain, one can always construct an $m$-site non-epistatic fitness landscape and choice of starting genotype such that the resulting mean fitness trajectory $f^{*}(t)$ satisfies
\begin{equation}
\sup_{t\geq 0} |f^{*}(t)-f(t)| \leq \frac{1}{m+1}\sqrt{\frac{\operatorname{Var}_{\boldsymbol{\pi}} \mathbf{F}}{\boldsymbol{\pi}(i)}}.
\end{equation}
While we make no claims as to the tightness of this bound, it is worth noting that the true error is often much lower than the bound would suggest, particularly if we only look at a finite range of times such as for the landscapes investigated in Table~\ref{tab:unconst}.

The derivation has two parts. First we show that we can construct a landscape such that the error is at most $\sum_{k=2}^{n}|c_{k}|/(m+1)$, where the $c_{k}$ are defined in Equation~\ref{eq:fitnessdecomp}. The proof is based on a closely related argument from~\citet[pg.~768]{Kammler76}, which the interested reader should also consult concerning the relation to completely monotone functions and the Laplace-Stieltjes transform. The second part of the proof then uses H\"{o}lder's inequality together with some linear algebra to bound $\sum_{k=2}^{n}|c_{k}|$ in terms of the equilibrium frequency of the initial genotype and the variance in fitness at equilibrium for the original epistatic fitness landscape.

It is sufficient to construct approximations to mean fitness trajectories of the form
\begin{equation}
f(t)=\sum_{k=2}^{n} c_{k} e^{-\lambda_{k}t} \quad \text{with }\lambda_{k}>0,
\end{equation}
where we have assumed without loss of generality that $c_{1}=0$ (we could match any value by appropriately choosing the initial fitness on the non-epistatic fitness landscape) and $c_{k} \neq 0$ for $k\geq 2$. Now, by our main result, we can construct an $m$-site non-epistatic fitness landscape that produces any mean fitness trajectory of the form
\begin{equation}
f^{*}(t)=\sum_{i=1}^{m'} c^{*}_{i} e^{-\lambda^{*}_{i}t} \quad \text{with }\lambda^{*}_{i}>0,
\end{equation}
where we choose $1\leq m'\leq m$. Furthermore, we can pick the $\lambda_{i}^{*}$ and $c_i^{*}$ such that, for any $\lambda\geq 0$, we have 
\begin{equation}
\left| \left(\sum_{i:\lambda_{i}^{*} \leq \lambda} c_{i}^{*} \right)-\left(\sum_{k:\lambda_{k} \leq \lambda} c_{k}\right) \right| \leq \frac{1}{m+1} \sum_{k} \left| c_{k} \right|.
\end{equation}
For instance, we can choose
\begin{equation}
\lambda_{i}^{*}=\sup \{\lambda> \lambda_{i-1}^{*}: \sum_{k: \lambda_{i-1}^{*} <\lambda_{k} \leq_{\lambda} }|c_{k}| \leq \frac{1}{m+1} \sum_{k} \left| c_{k}\right| \}
\end{equation}
\begin{equation}
c_{i}^{*}=\sum_{k:\lambda_{i-1}^{*}<\lambda_{k} \leq \lambda^{*}_{i}} c_{k},
\end{equation}
where we interpret $\lambda_{0}^{*}$ as $0$ and define the $\lambda_{i}^{*}$ iteratively for $i=1,2,\ldots$ until we either reach $m$ or the $\sup$ no longer exists in which case we set $m'$ equal to the last value of $i$ for which the $\sup$ exists. To see why this solution works, note that for $\lambda \in \{\lambda_{1}^{*},\ldots, \lambda_{m'}^{*} \}$
\begin{equation}
\left| \left(\sum_{i:\lambda_{i}^{*} \leq \lambda} c_{i}^{*} \right)-\left(\sum_{k:\lambda_{k} \leq \lambda} c_{k}\right) \right|=0
\end{equation}
and that the sum $\sum_{k:\lambda_{k} \leq \lambda} c_{k}$ viewed as a function of $\lambda$ can change its value by at most $\sum_{k}|c_{k}|/(m+1)$  in each of the intervals $[0,\lambda_{1}^{*}),[\lambda_{1}^{*},\lambda_{2}^{*}),\ldots, [\lambda_{m'}^{*},\infty)$.

Having specified our approximating mean fitness trajectory $f^{*}(t)$, we can now bound its error relative to $f(t)$. Note that for any $x,t>0$, we can write $e^{-xt}=t \int_{x}^{\infty}e^{-\lambda t}\, d\lambda= t \int_{0}^{\infty}\chi(\lambda-x)\, e^{-\lambda t} \,d \lambda$, where $\chi(y)=1$ for $y\geq 0$ and $0$ otherwise. Thus we have, for $t>0$:
\begin{align}
\left| f^{*}(t)-f(t) \right| &= \left| \left(\sum_{i=1}^{m'} c^{*}_{i} e^{-\lambda_{i}^{*} t} \right) -\left(\sum_{k=1}^{m} c_{k} e^{-\lambda_{k} t} \right)\right| \\
 &= \left| \left(\sum_{i=1}^{m'} c^{*}_{i}\, t \int_{0}^{\infty} \chi(\lambda-\lambda^{*}_{i})\, e^{-\lambda t}\,d\lambda\right) -\left(\sum_{k=1}^{m} c_{k}\, t \int_{0}^{\infty}  \chi(\lambda-\lambda_{k})\,  e^{-\lambda t}\, d\lambda\right)\right| \\
 &= \left| t  \int_{0}^{\infty} \left( \left(\sum_{i=1}^{m'} c_{i}^{*}\, \chi(\lambda-\lambda^{*}_{i})\right)-\left(\sum_{k=1}^{m} c_{k}\, \chi(\lambda-\lambda_{k})\right) \right) e^{-\lambda t}\,d\lambda \right| \\
 &= \left| t \int_{0}^{\infty} \left( \left( \sum_{i:\lambda_{i}^{*} \leq \lambda} c_{i}^{*} \right) -\left(\sum_{k:\lambda_{k} \leq \lambda} c_{k} \right) \right) e^{-\lambda t}\, d\lambda \,\right| \\
 & \leq  t \int_{0}^{\infty} \left| \left( \sum_{i:\lambda_{i}^{*} \leq \lambda} c_{i}^{*} \right) -\left(\sum_{k:\lambda_{k} \leq \lambda} c_{k} \right) \right| e^{-\lambda t}\, d\lambda \\
 & \leq \left(\sup_{\lambda} \left| \left( \sum_{i:\lambda_{i}^{*} \leq \lambda} c_{i}^{*} \right) -\left(\sum_{k:\lambda_{k} \leq \lambda} c_{k} \right) \right| \right)t \int_{0}^{\infty} e^{-\lambda t}\, d\lambda \\
 & = \sup_{\lambda} \left| \left( \sum_{i:\lambda_{i}^{*} \leq \lambda} c_{i}^{*} \right) -\left(\sum_{k:\lambda_{k} \leq \lambda} c_{k} \right) \right| \\
 & \leq \frac{1}{m+1} \sum_{k} \left| c_{k} \right|.
\end{align}
This establishes the required inequality for $t>0$; the inequality must then also hold at $t=0$ by the continuity of $f(t)$ and $f^{*}(t)$.

It remains to derive an upper bound on $\sum_{k}|c_{k}|=\sum_{k=2}^{n} \mathbf{r}_{k} (i)\, \transpose{\mathbf{l}_{k}} \mathbf{F}$. By H\"{o}lder's inequality we have 
\begin{equation}
\sum_{k=2}^{n}\mathbf{r}_{k} (i)\, \transpose{\mathbf{l}_{k}} \mathbf{F} \leq \sqrt{\sum_{k=2}^{n}\left(\mathbf{r}_{k} (i)\right)^{2} } \sqrt{\sum_{k=2}^{n}\left(\transpose{\mathbf{l}_{k}} \mathbf{F} \right)^{2}}.
\end{equation}
Now, $\sum_{k=2}^{n}\left(\mathbf{r}_{k} (i)\right)^{2}  \leq \sum_{k=1}^{n}\left(\mathbf{r}_{k} (i)\right)^{2}$ and the latter sum is the squared Euclidean norm of the $i$-th row of the matrix $\mathbf{D}_{\boldsymbol{\pi}}^{-1/2} \mathbf{U}$, where $\mathbf{U}$ is the matrix with $\mathbf{u}_{k}$ as its $k$-th column. Since $\mathbf{U}$ is an orthogonal matrix, its rows are orthonormal and hence have a squared Euclidean norm equal to 1. Because the $i$-th row of $\mathbf{U}$ is multiplied by $1/\sqrt{\boldsymbol{\pi}(i)}$ in the matrix product $\mathbf{D}_{\boldsymbol{\pi}}^{-1/2} \mathbf{U}$, we have $\sum_{k=1}^{n}\left(\mathbf{r}_{k} (i)\right)^{2}=1/\boldsymbol{\pi}(i)$. Indeed, since $\mathbf{r}_{1}(i)=1$ for all $i$, we have 
\begin{equation}
\sum_{k=2}^{n}\left(\mathbf{r}_{k} (i)\right)^{2} = \frac{1-\boldsymbol{\pi}(i)}{\boldsymbol{\pi}(i)} \leq \frac{1}{\boldsymbol{\pi}(i)}.
\end{equation}

As for the other sum, since $\mathbf{l}_{1}=\boldsymbol{\pi}$, we have
\begin{align}
\sum_{k=2}^{n}\left(\transpose{\mathbf{l}_{k}} \mathbf{F} \right)^{2} &= \left( \sum_{k=1}^{n}\left(\transpose{\mathbf{l}_{k}} \mathbf{F} \right)^{2} \right) - \left(\transpose{ \boldsymbol{\pi}}\mathbf{F} \right)^{2}\\
 & =  \left( \sum_{k=1}^{n}\left(\transpose{ \left( \mathbf{D}_{\boldsymbol{\pi}}^{1/2}\mathbf{u}_{k} \right)} \mathbf{F} \right)^{2} \right) - \left( \transpose{ \boldsymbol{\pi} }\mathbf{F} \right)^{2} \\
 & =  \left( \sum_{k=1}^{n}\left(\transpose{  \mathbf{u}_{k}} \left(\mathbf{D}_{\boldsymbol{\pi}}^{1/2} \mathbf{F} \right) \right)^{2} \right) - \left( \transpose{ \boldsymbol{\pi} }\mathbf{F} \right)^{2} \\
 &= \left( \sum_{k=1}^{n}\left(\sqrt{\boldsymbol{\pi}(k)}\mathbf{F}(k) \right)^{2} \right) - \left( \transpose{ \boldsymbol{\pi} }\mathbf{F} \right)^{2} \\
 & =  \transpose{ \boldsymbol{\pi} }\mathbf{F}^{2} -\left( \transpose{ \boldsymbol{\pi} }\mathbf{F} \right)^{2} \\
 & = \operatorname{Var}_{\boldsymbol{\pi}} \mathbf{F},
\end{align}
where $\mathbf{F}^{2}(i)=\mathbf{F}(i)^{2}$ and we have used the fact that $\mathbf{U}$ is orthonormal and hence preserves the squared Euclidean norm of a vector. This completes the derivation of the bound.

To study evolution at equilibrium, we again consider an ensemble of populations, but instead of assuming that all populations in the ensemble begin at some specified genotype, we let the initial genotype of each population be drawn from $\boldsymbol{\pi}$, the equilibrium distribution of the Markov chain defined by $\mathbf{Q}$.  Using the definition of covariance, the covariance between the fitness of a population at time $t'\geq0$ whose genotype is drawn from $\boldsymbol{\pi}$ at time $0$ and its fitness at time $t'+t$ is given by
\begin{equation}
a(t)=\sum_{i=1}^{n}\sum_{j=1}^{n}\boldsymbol{\pi}(i)\,\mathbf{P}_{t}(i,j) \left( \mathbf{F}(i)-\transpose{\boldsymbol{\pi}}\mathbf{F} \right)\left( \mathbf{F}(j)-\transpose{\boldsymbol{\pi}}\mathbf{F} \right).
\end{equation}
Defining the centered fitness vector $\mathbf{F}'=\mathbf{F}-(\transpose{\boldsymbol{\pi}} \mathbf{F}) \mathbf{1}$, we can rewrite this in matrix notation as
\begin{equation}
a(t)=\transpose{(\mathbf{F}')} \mathbf{D}_{\boldsymbol{\pi}} \mathbf{P}_{t}\mathbf{F}'.
\end{equation}
Using Equation~\ref{eq:eigendecomp}, we can then expand $\mathbf{P}_{t}$ in terms of its eigenvalues and eigenvectors and simplify to get
\begin{align}
a(t)&=\sum_{k=1}^{n} e^{-\lambda_{k}t}\left(\transpose{(\mathbf{F}')} \mathbf{D}_{\boldsymbol{\pi}} \mathbf{r}_{k} \right)\left(\transpose{\mathbf{l}_{k}} \mathbf{F}' \right)\\
 &=\sum_{k=1}^{n} e^{-\lambda_{k}t}\left(\transpose{\mathbf{l}_{k}} \mathbf{F}' \right)^{2} \\
 &=\sum_{k=2}^{n} e^{-\lambda_{k}t}\left(\transpose{\mathbf{l}_{k}} \mathbf{F} \right)^{2} \label{eq:squared}
\end{align}
where the last line follows because by construction $\transpose{\mathbf{l}_{1}} \mathbf{F}'=\transpose{\boldsymbol{\pi}} \mathbf{F}'=0$ and, for $k\geq 2$, $\transpose{\mathbf{l}_{k}} \mathbf{1}=\transpose{\mathbf{l}_{k}} \mathbf{r}_{1}=0$, so that for $k\geq 2$
\begin{align}
\transpose{\mathbf{l}_{k}} \mathbf{F}' &=\transpose{\mathbf{l}_{k}} \mathbf{F}-\left(\transpose{\boldsymbol{\pi}} \mathbf{F}\right)\left( \transpose{\mathbf{l}_{k}} \mathbf{1} \right)\\
 &= \transpose{\mathbf{l}_{k}} \mathbf{F}.
\end{align}
Equation~\ref{eq:autoco} in the main text then follows from Equation~\ref{eq:squared} by noting that $\left(\transpose{\mathbf{l}_{k}} \mathbf{F} \right)^{2}$ is non-negative.

\end{document}